\documentclass{aastex}
\usepackage{spr-astr-addons}
\usepackage[colorlinks,linkcolor=cyan,citecolor=blue]{hyperref}

\graphicspath{{./figs/}}   
\begin{document}

\title{Estimating Red Noise in Quasi-periodic Signals with MCMC-Based Bayesian}
\slugcomment{Not to appear in Nonlearned J., 45.}
\shorttitle{Estimating Red Noise with MCMC-Based Bayesian}
\shortauthors{B. LIANG et al.}

\author {Bo LIANG\altaffilmark{1,2,3}}
\and
\author{Yao MENG\altaffilmark{1,3}}
\and
\author{Song FENG\altaffilmark{1,2}}
\and
\author{Yunfei YANG\altaffilmark{1,3}}

\altaffiltext{1}{Faculty of Information Engineering and Automation, Kunming University of Science and Technology, Kunming 650500, China. feng.song@kust.edu.cn}
\altaffiltext{2}{Key Laboratory of Solar Activity, National Astronomical Observatories, Chinese Academy of Sciences, Beijing 100012, China}
\altaffiltext{3}{Yunnan Key Laboratory of Computer Technology Application, Kunming 650500, China.}

\begin{abstract}
Multi-parameter Bayesian inferences based on Markov chain Monte Carlo (MCMC) samples have been widely used to estimate red noise in solar period-periodic signals. 
To MCMC, proper priors and sufficient iterations are prerequisites ensuring the accuracy of red noise estimation. 
We used MCMC-based Bayesian inferences to estimate 100 groups of red noise synthesized randomly for evaluating its accuracy. 
At the same time, the Brooks-Gelman algorithm was employed to precisely diagnose the convergence of the Markov chains generated by MCMC. 
The root-mean-square error of parameter inferences to the synthetic data is only 1.14.
Furthermore, we applied the algorithm to analyze the oscillation modes in a sunspot and a flare. 
A 70 s period is detected in the sunspot umbra in addition to 3- and 5-minute periods, and a 40 s period is detected in the flare.
The results prove that estimating red noise with MCMC-based Bayesian  has more high accuracy in the case of proper priors and convergence. 
We also find that the number of iterations increases dramatically to achieve convergence as the number of parameters grows. 
Therefore, we strongly recommend that when estimating red noise with MCMC-based Bayesian, different initial values must be selected to ensure that the entire posterior distribution is covered.

\end{abstract}

\keywords{method: data analysis --- method: statistical --- Sun: oscillations --- Sun: sunspots --- Sun: flares}

\section{Introduction}
\label{intr}
 Solar quasi-periodic oscillations are rhythmic modulations to electromagnetic radiation of plasma in the solar atmosphere.
The oscillations are usually observed in integrated light curves from radio waveband, optical, extreme ultraviolet to X-rays. 
For example, quasi-periodic pulsation (QPP) in solar flares \citep{2015ApJ...813...59L,2018ApJ...868L..33L,2017SoPh..292...11N,2016ApJS..224...30Y}, three- and five-minute oscillations in sunspots \citep{2012A&A...539L...4S,2016ApJ...816...30S,2018ApJ...856L..16W,2014ApJ...786..137T,2014ApJ...792...41Y,2020RAA....20....6W}.

It is well known that the oscillation signals are usually suppressed by red noise in observations \citep{2011A&A...533A..61G,2015ApJ...798..108I,2017SoPh..292...11N,2017A&A...602A..47P,2019ApJ...886L..25Y}. 
Accurate estimating red noise in a light curve is crucial to extract and represent the solar oscillation modes.
Power spectral densities (PSDs) of oscillation signals affected by red noise present a power law distribution, i.e., $P(f)=Af^{-\alpha}$.
In solar physics, traditional practice is to use Bayesian inference based Markov chain Monte Carlo (MCMC-based Bayesian) to estimate the significance of those peaks in frequency domain \citep{2010MNRAS.402..307V,2011A&A...533A..61G,2015ApJ...798..108I,2016ApJ...833..284I,2017A&A...597L...4L,2019ApJ...886L..25Y}.

To parameter estimation with the MCMC-based Bayesian inference, two important practical issues are convergence and proper priors.
The irreducible, aperiodic, and recurrent characteristic is essential to the convergence of a Markov chain.  
A random walk may guarantee the aperiodicity and the recurrence.
So, we must perform numerous samples to convince us that the Markov chain has covered the entire posterior distribution and reached convergence.
In the procedure drawing samples, some samples obtained during the burn-in period must also be discarded for subsequent analysis because they do not represent the given probability density function. 
The other issue is that improper priors or lack of prior knowledge may cause incorrect results. 
Thus, the convergence of the Markov chain generated by MCMC-based Bayesian and the proper priors are very important to estimate red noise in a quasi-periodic signal.

In the paper, we assessed the accuracy using MCMC-based Bayesian to estimate red noise and also diagnosed its convergence in the case of proper priors and sufficient iterations. 
The structure of this paper is as follows: Section \ref{algorithm} introduces the MCMC algorithm and the convergence diagnostics algorithm.
Section \ref{bayesian} assesses the accuracy with the MCMC-based Bayesian, and
Section \ref{result} uses the method to extract the oscillation modes in a sunspot and a flare.
Finally, Section \ref{conclusion} discusses our results and concludes our study.

\section{Algorithm}
\label{algorithm}

\subsection{Metropolis-Hastings Algorithm }
To MCMC, Metropolis-Hastings (MH) algorithm is a common practices  \citep{1953JChPh..21.1087M,1970Bimka..57...97H}. 
It draws samples from a given distribution up to a constant. 
Random numbers are generated from the distribution with a probability density function that is equal to or proportional to a proposal distribution.
To generate the random numbers, an initial random value $x(t)$ is given, and then to draw a sample $y(t)$ from a prior distribution $q(y|x)$.   $y(t)$ is accepted as the next sample $x(t + 1)$ with probability $\alpha(x,y)$, or $x(t)$ is kept as the next sample  $x(t + 1)$ with probability $1 - \alpha(x,y)$, where:
\begin{equation}
\alpha(x,y)=\min \left\{\frac{\pi(y) q(x | y)}{\pi(x) q(y | x)}, 1\right\}.
\end{equation}

Repeating the process until reach the desired number of samples.
More detailed descriptions about the algorithm can be found in \cite{1970Bimka..57...97H}. 
\begin{figure*}[t]
	\centering
	\includegraphics[width=16cm]{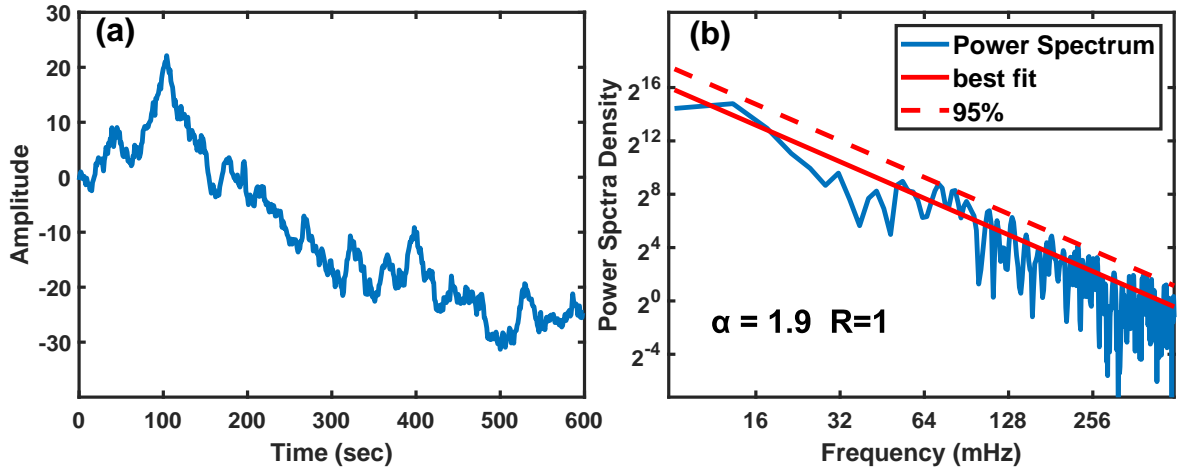}
	\caption{
		A red noise (panel a) synthesized and its PSD (panel b) in log-log space. 
		The curve of red noise (panel a) shows aperiodic, erratic fluctuations. 
		 The theoretical value of the PSD (blue) is 2 and the fitted value is 1.9.
		The dashed line indicates the 95\% confidence level (i.e., 5\% significance level). We note that there is no peak above the confidence level.}
	\label{fig1}
\end{figure*}

\subsection{Convergence Diagnostics}
\label{Compar}
 We used the Brooks-Gelman algorithm \citep{Brooks:1998ci} to diagnose the convergence of a Markov chain generated by MH.
The Brooks-Gelman algorithm is based on the hypothesis that if multiple Markov chains have converged, the chains should appear very similar to each other.
A convergence factor $\widehat{R}$ is defined to diagnose the similarity of Markov chains.
The convergence diagnostics has $m$ chains with $2n$ iterations but be used only the last $n$ iterations for removing the effect of the starting distribution.
So, $\widehat{R}$ is defined with between-chain variance $B$ and within-chain variance $W$. Assuming $m$ chains, each of length $n$, quantities as calculated by
\begin{equation}
\left\{
\begin{aligned}
B &=\frac{1}{m - 1} \sum_{j = 1}^{m}\left(\bar{\theta}_{j} - \bar{\theta}\right)^{2} \\
W &=\frac{1}{m} \sum_{j = 1}^{m}\left[\frac{1}{n - 1} \sum_{i = 1}^{n}\left(\theta_{i j} - \bar{\theta}_{, j}\right)^{2}\right]
\end{aligned}
\right.
\end{equation}
for each parameter $\theta$. Using these values, we estimate the posterior variance of $\theta$:
\begin{equation}
\widehat{V}=\frac{n - 1}{n} W + \frac{m+1}{m} B.
\end{equation}
So, the convergence factor is calculated by
\begin{equation}
\widehat{R}=\sqrt{\frac{d+3}{d+1} \frac{\widehat{V}}{W}},
\end{equation}
where, $d$ is the degree of freedom, and can be estimated by $d \approx 2\widehat{V}/ {Var(\widehat{V})}$.
$\widehat{R}$ above 1 indicates lack of convergence.

\section{Red Noise Estimation}
\label{bayesian}
We synthesized 100 groups of red noise data. 
Each red noise is composed of 600 data points, and each sample interval is 1 second with $\alpha$\ equal to 2 power spectral shape.
Figures \ref{fig1}a and b show a red noise datum and its PSD in log-log space, respectively.
The red noise exhibits aperiodic, erratic fluctuations, and its PSD presents a decay linear trend.

Subsequently, we used the MCMC-based Bayesian method to infer the parameter values of the red noise, i.e., $\alpha$ and $A$.
Bayesian inference is to find the probability distribution of the parameters of the given model $p(\theta|D, M)$, where $D$ denotes the PSD of red noise, and $M$ the PSD of the model, $P=Af^{-\alpha}$. 
The parameter $\theta$ probability may be obtained in terms of Bayesian theorem,
\begin{equation}
p(\theta|D, M)=\frac{p(D| \theta, M) p(\theta|M)}{p(D|M)},
\end{equation}
where $p(\theta|M)$ is a prior function.
$p(D|\theta, M)$  is a likelihood function, i.e.,  the probability of assuming data $D$ under model $M$.
The denominator $p(D|M)$ is a normalization constant.
Therefore, posterior probability $p(\theta|D, M)$ is described by the product of  $p(\theta|M)$ and $p(D| \theta, M)$.

To a random time-series of length $N$, its PSD $D_j$ at Fourier frequency $f_j$, is exponentially distributed with the real PSD $M_j$.
So, the likelihood function is defined, in term of Nyquist frequency, as \citep{chatfield2003analysis}
\begin{equation}
p(D|\theta,M) = \prod_{j=1}^{N/2}\frac{1}{M_j}exp(-\frac{D_j}{M_j}).
\end{equation}

\begin{figure*}[t]
	\centering
	\includegraphics[width=12cm]{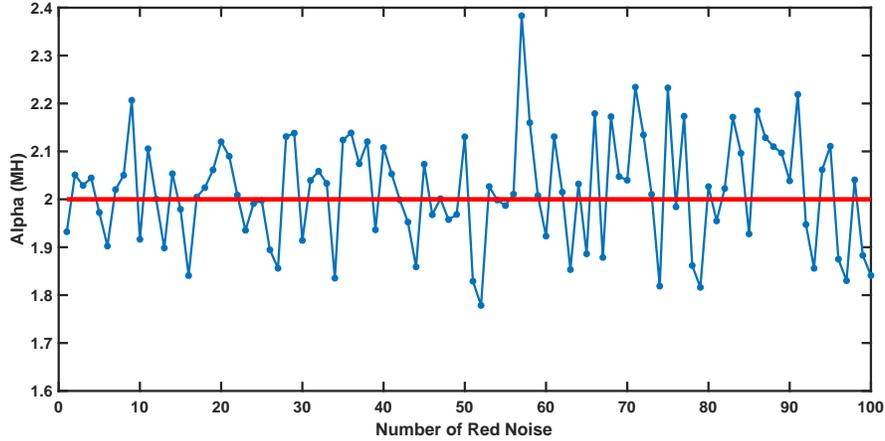}
	\caption{
		Result of parameter estimation to $\alpha$. 
		The red line indicates the theoretical values of parameter $\alpha$ that are all set to 2. 
		The asterisks mark the estimated result to $\alpha$ of each red noise. 
		The root mean squared error between the estimated and theoretical values is 1.14.}
	\label{fig2}
\end{figure*}
\begin{figure*}[t]
	\centering
	\includegraphics[width=12cm]{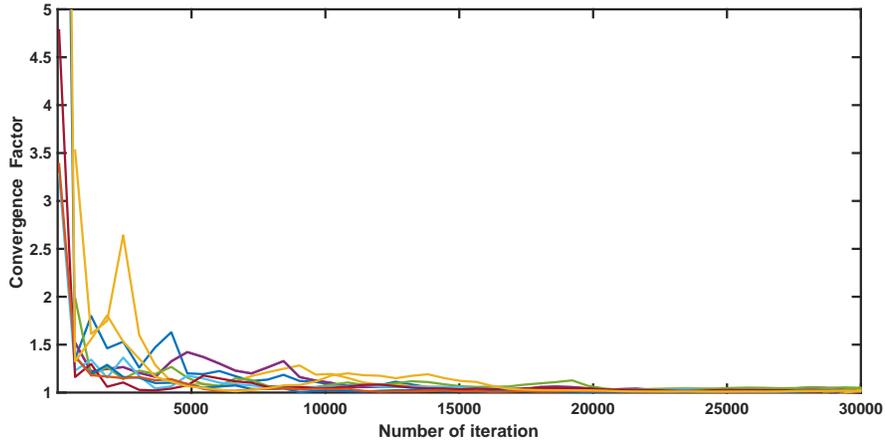}
	\caption{
		Brooks-Gelman convergence factor $\widehat{R}$ based on 4 chains. 
		Here, we only show the change of the convergence factor of the 10 groups of red-noise data that are marked with different colors. 
		The convergence factor $\widehat{R}$ approaches unity when the number of iterations reaches 25,000. 
		This implies that the Markov chains have reached convergence.
	}
	\label{fig3}
\end{figure*}

\begin{figure*}[t]
	\centering
	\includegraphics[width=16cm]{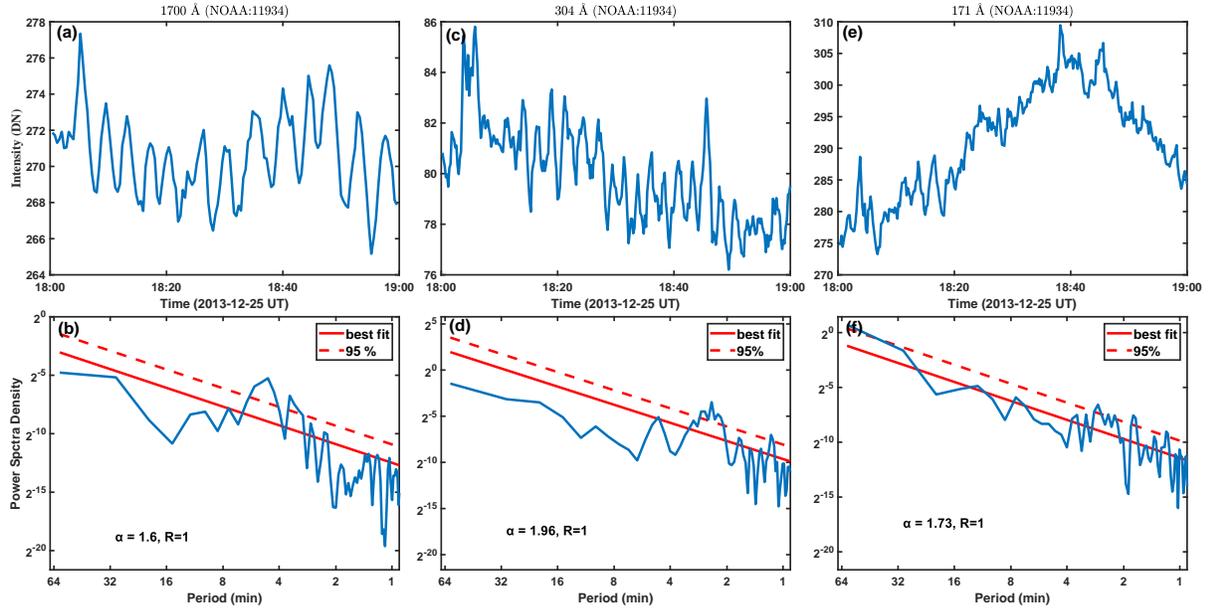}
	\caption{
		Red noise estimation to umbral data observed by SDO/AIA  on 2013 December 25 that located in active region 11934.
		The light curves of three channels, 1700 \AA, 304 \AA, and 171 \AA\, are shown in the upper row, and their corresponding PSDs in log-space are exhibited in the bottom row.
	}
	\label{fig4}
\end{figure*}
\begin{figure*}[t]
	\centering
	\includegraphics[width=16cm]{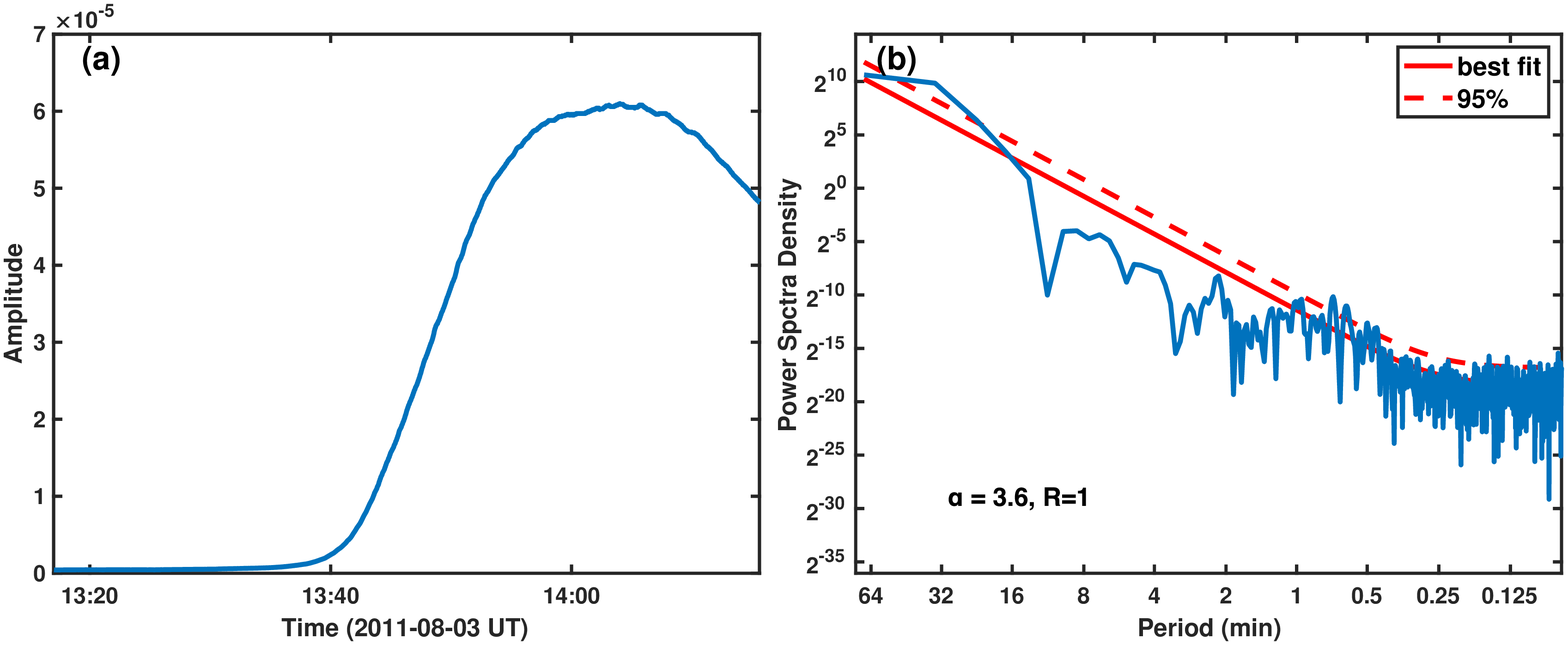}
	\caption{Red noise estimation to a flare observed by GOES on 2011 August 3. (a):Soft X-ray flux of the flare. b: PSD of the flux.}
	\label{fig5}
\end{figure*}

In log-log space, since each parameter of the red noise model in log-log space is affected by each independent frequency component, parameters $\alpha$ and $log(A)$ follow independent normal distributions.
\cite{2010MNRAS.402..307V} also proved that the fit result is optimum if the priors of the two parameters are set to independent normal distributions.
Here, $\alpha$ ranges from 0 to 4, and $log(A)$ ranges from -20 to 0.

In Figure \ref{fig1}b, the best fit  and the corresponding 95\% confidence level (i.e., 5\% significance level) are marked with red solid and dashed lines, respectively. Here,  the slope of the best fit line (i.e., $\alpha$) is equal to 1.9.
The PSD of red noise follows a chi-square distribution due to the real and the imaginary follow independent normal distributions. Thus, we used a chi-square test to assess the significance of the peaks in the PSD.
The chi-square test has also been used \citep{2005A&A...431..391V,2010MNRAS.402..307V,2015ApJ...798..108I}  to infer the significance of red noise.
The peaks above the confidence level are considered as frequency components instead of red noise.
There are no peaks above 95\% confidence level in Figure \ref{fig1}b, meaning that the curve shown in Figure \ref{fig1}a is noise instead of a signal. 
The result demonstrates that estimating red  noise with MCMC-based Bayesian holds a quite high accuracy.

Figure \ref{fig2} shows the inference results to the 100 groups of red noise data.
The red line indicates the position of the theoretical value of parameter $\alpha$, and every asterisk marks an estimation value to $\alpha$. 
Here, the number $m$ of the Markov chains is equal to 4, the iteration number set to 45,000, and burn-in 15,000.
The estimated values fluctuate in the range between 1.7 and 2.3. 
The root-mean-square error (RMSE) between estimated and theoretical values  is 1.14, demonstrating that the parameter estimation with MCMC-based Bayesian is very accurate. 
The $\alpha$ values are obtained when the convergence factor $\widehat{R}$ is equal to unity. 

Figure \ref{fig3} illustrates $\widehat{R}$ changes as the number of iteration increases. 
Here, we only illustrate 10 groups of red noise due to similar trends. 
The different colors indicate the convergence process of different red noise data.
We can find that the Markov chains reach convergence more than 25,000 iterations.

\section{Red Noise Estimation to Solar signals}
\label{result}
\subsection{Sunspot}
Figure \ref{fig4} show three umbral light curves observed by the Atmospheric Imaging Assembly \citep[AIA,][]{{2012SoPh..275...41B}} onboard the Solar Dynamics Observatory \citep[SDO,][]{2012SoPh..275....3P} in three channels, 1700 \AA, 304 \AA, and 171 \AA, and their PSDs.
The sunspot was captured on 2013 December 25 starting from 18:00 UT for 60 minutes that located in active region 11934. 

The three PSDs shown in the bottom row of Figure \ref{fig4} present power law distributions. 
So we selected the noise model, $S_1=Af^{-\alpha}$, and defined the priors of parameter $\alpha$ and $A$ as given in Section \ref{bayesian}.
The best fits and their 95\% confidence levels are marked with red solid and dashed lines, respectively. Here, to the three channels, the $\alpha$ values are 1.6, 1.96, and 1.73 when $\widehat{R}$ equal to unity, respectively.

Those peaks above the confidence levels in the bottom row of Figure \ref{fig4} are considered as periodic components. 
We note two significant peaks with  4.6 min and 3.5 min periodicities in Figure \ref{fig4}b. 
In Figure \ref{fig4}d, there are three significant peaks whose periodicities are 2.2 min, 2.4 min, and 2.7 min. 
Interestingly, the three same periodicities above the confidence level in 171 \AA\ channel are also detected. 
But, the 4.6 min and 3.5 min periodicities in 1700 \AA\ channel are not detected in 304 \AA\ and 171 \AA\ channels.
We also note that in 304 \AA\ and 171 \AA\ channels, a significant peak with a 70 s periodicity.  
The periods with 1--2 minutes are also been found in an umbra \citep[see e.g., ][]{2018ApJ...856L..16W}.

\subsection{Solar Flare}

We selected a class M 6.0 flare detected by the Geostationary Operational Environmental Satellite (GOES) on 2011 August 3 on active region 11520. 
Figures \ref{fig5}a and b present the Soft X-ray flux and its PSD. 
The Soft X-ray flux started to raise at about 13:17 UT, reached a peak value at about 13:48 UT and decays off gradually. 

In Figure \ref{fig5}b, The PSD exhibits a power law distribution before the 0.25 minute period in the X-axis direction, and next presents a flat spectrum. 
This implies that the high-frequency components of the flare are affected by white noise in addition to red noise. 
In this case, the noise model is defined as $S_2=Af^{-\alpha}+C$ with three parameters $\theta=\{A,\alpha, C\}$. 
$C$ denotes white noise.  
Similar to Figure \ref{fig4}, the best fit and its 95\% confidence level are marked with red solid and dashed lines, respectively. 
Due to add parameter $C$ in the noise model, the Markov chain can achieve convergence more than 30,000 iterations. 
There are several peaks above the 95 \% confidence level near 40 s, indicating that the flare exists a period with about 40 s.
The period has been reported \citep[see e.g.,][]{2017MNRAS.471L...6L,2017A&A...597L...4L}.

\section{Discussion and Conclusion}
\label{conclusion}

At present, MCMC-based Bayesian has been widely used to estimate red noise in quasi-periodic signals \citep[see e.g., ][]{2015ApJ...798..108I,2010MNRAS.402..307V}.  
MCMC is a method drawing samples  from a distribution, and Bayesian is a theory to interpret observed data. 
But, the MCMC-based Bayesian inference has two important issues, i.e., convergence and proper priors.
Improper prior knowledge or insufficient MCMC iterations may cause incorrect results.
So, we often have to face the question whether the inference obtained by MCMC-based Bayesian is correct.
Most previous studies mainly focus on using the method to estimate red noise.
Therefore, we give attention to the accuracy of noise estimation and its convergence when using MCMC-based Bayesian to estimate the red noise in a quasi-periodic signal.

Due to red noise follows a power law distribution, the red noise spectrum model is defined as $P(f)=Af^{-\alpha}$. Parameter $\alpha$ prior is defined by a normal distribution and $A$ follows a log-normal distribution.

Figure \ref{fig1}a shows a red noise curve and its spectrum. The best fit and its 95\% confidence level are plotted in Figure \ref{fig1}b. The fit result is obtained after 30,000 iterations and the chain reaches convergence. Figure \ref{fig2} exhibits the estimation result to 100 groups of red noise in the case of the same priors and the number of iterations. The result that no peaks above the 95\% confidence level in Figure \ref{fig1}b and the RMSE of parameter inference is 1.14 in Figure \ref{fig2} proves that estimating red noise with MCMC-based Bayesian hold a high accuracy in the case of proper priors and sufficient iterations.

We further analyzed the red noise in the SDO/AIA and GOES observations and found that, to the sunspot, the number of iterations still takes 25,000 times, but to the flare, it would require at least 30,000 and reaches convergence.

In order to evaluate the effect of the number of iterations on the convergence, we selected two more complex models $S_3=S_1 + D$ and $S_4=S_2 + D$ proposed by \citep{2015ApJ...798..108I} to again fit red noise in the sunspot and flare data. 
Where $D=H e^{-(log f - \mu)^2/2\sigma^2}$ denotes a Gaussian function to estimate the position of oscillation modes. 
$H$, $\mu$, and $\sigma$ denote height, position, and  spectral width of the oscillation modes, respectively. 
In this case, models $S_3$ and $S_4$ contain 5 and 6 parameters, respectively.
We found that model $S_3$ requires about 60,000 iterations, while $S_4$ about 80,000 iterations to reach convergence.
This implies that  the number of iterations increases as the number of parameters grows.

To our knowledge, we first prove that the accuracy of MCMC-based Bayesian inference when estimating red noise in quasi-periodic signals. 
However, we also find that insufficient iterations could cause divergence of Markov chains generated by MCMC even though the priors are proper.
Therefore, we strongly recommend that the convergence should be performed, and meanwhile, different initial values should also be given to guarantee that they cover the entire posterior distribution. 
This is an essential prerequisite to ensure the accuracy of red noise estimation.

\acknowledgments
We would like to thank the anonymous referee for helpful comments. 
S. Feng is supported by the Joint Funds of the National Natural Science Foundation of China (U1931107), the Key Applied Basic Research Program of Yunnan Province (2018FA035), and the Open Research Program (KLSA202007) of Key Laboratory of Solar Activity of National Astronomical Observatory of China. Y. Yang is supported by NSFC (11763004).
We thank the  science teams of the GOES, SDO/AIA.


\begin{thebibliography}{25}
	\ifx \bisbn   \undefined \def \bisbn  #1{ISBN #1}\fi
	\ifx \binits  \undefined \def \binits#1{#1} \fi
	\ifx \bauthor  \undefined \def \bauthor#1{#1} \fi
	\ifx \batitle  \undefined \def \batitle#1{#1} \fi
	\ifx \bjtitle  \undefined \def \bjtitle#1{#1}\fi
	\ifx \bvolume  \undefined \def \bvolume#1{\textbf{#1}}\fi
	\ifx \byear  \undefined \def \byear#1{#1} \fi
	\ifx \bissue  \undefined \def \bissue#1{#1} \fi
	\ifx \bfpage  \undefined \def \bfpage#1{#1} \fi
	\ifx \blpage  \undefined \def \blpage #1{#1} \fi
	\ifx \burl  \undefined \def \burl#1{\textsf{#1}} \fi
	\ifx \doiurl  \undefined \def \doiurl#1{\textsf{#1}} \fi
	\ifx \betal  \undefined \def \betal{\textit{et al.}} \fi
	\ifx \binstitute  \undefined \def \binstitute#1{#1} \fi
	\ifx \binstitutionaled  \undefined \def \binstitutionaled#1{#1} \fi
	\ifx \bctitle  \undefined \def \bctitle#1{#1} \fi
	\ifx \beditor  \undefined \def \beditor#1{#1} \fi
	\ifx \bpublisher  \undefined \def \bpublisher#1{#1} \fi
	\ifx \bbtitle  \undefined \def \bbtitle#1{#1} \fi
	\ifx \bedition  \undefined \def \bedition#1{#1} \fi
	\ifx \bseriesno  \undefined \def \bseriesno#1{#1} \fi
	\ifx \blocation  \undefined \def \blocation#1{#1} \fi
	\ifx \bsertitle  \undefined \def \bsertitle#1{#1} \fi
	\ifx \bsnm \undefined \def \bsnm#1{#1} \fi
	\ifx \bsuffix \undefined \def \bsuffix#1{#1} \fi
	\ifx \bparticle \undefined \def \bparticle#1{#1} \fi
	\ifx \barticle \undefined \def \barticle#1{#1} \fi
	\ifx \bconfdate \undefined \def \bconfdate #1{#1} \fi
	\ifx \botherref \undefined \def \botherref #1{#1} \fi
	\ifx \url \undefined \def \url#1{\textsf{#1}} \fi
	\ifx \bchapter \undefined \def \bchapter#1{#1} \fi
	\ifx \bbook \undefined \def \bbook#1{#1} \fi
	\ifx \bcomment \undefined \def \bcomment#1{#1} \fi
	\ifx \oauthor \undefined \def \oauthor#1{#1} \fi
	\ifx \citeauthoryear \undefined \def \citeauthoryear#1{#1} \fi
	\ifx \endbibitem  \undefined \def \endbibitem {}\fi
	\ifx \bconflocation  \undefined \def \bconflocation#1{#1} \fi
	\ifx \arxivurl  \undefined \def \arxivurl#1{\textsf{#1}} \fi
	
	\bibitem[\protect\citeauthoryear{Boerner et~al.}{2012}]{2012SoPh..275...41B}
	\begin{barticle}
		\bauthor{\bsnm{Boerner}, \binits{P.}},
		\bauthor{\bsnm{Edwards}, \binits{C.}},
		\bauthor{\bsnm{Lemen}, \binits{J.}},
		\bauthor{\bsnm{Rausch}, \binits{A.}},
		\bauthor{\bsnm{Schrijver}, \binits{C.}},
		\bauthor{\bsnm{Shine}, \binits{R.}},
		\bauthor{\bsnm{Shing}, \binits{L.}},
		\bauthor{\bsnm{Stern}, \binits{R.}},
		\bauthor{\bsnm{Tarbell}, \binits{T.}},
		\bauthor{\bsnm{Title}, \binits{A.}},
		\bauthor{\bsnm{Wolfson}, \binits{C.J.}},
		\bauthor{\bsnm{Soufli}, \binits{R.}},
		\bauthor{\bsnm{Spiller}, \binits{E.}},
		\bauthor{\bsnm{Gullikson}, \binits{E.}},
		\bauthor{\bsnm{McKenzie}, \binits{D.}},
		\bauthor{\bsnm{Windt}, \binits{D.}},
		\bauthor{\bsnm{Golub}, \binits{L.}},
		\bauthor{\bsnm{Podgorski}, \binits{W.}},
		\bauthor{\bsnm{Testa}, \binits{P.}},
		\bauthor{\bsnm{Weber}, \binits{M.}}:
		\bjtitle{Solar Physics}
		\bvolume{275}(\bissue{1}),
		\bfpage{41}
		(\byear{2012})
	\end{barticle}
	\endbibitem
	
	\bibitem[\protect\citeauthoryear{Brooks and Gelman}{1998}]{Brooks:1998ci}
	\begin{barticle}
		\bauthor{\bsnm{Brooks}, \binits{S.P.}},
		\bauthor{\bsnm{Gelman}, \binits{A.}}:
		\bjtitle{Journal of Computational and Graphical Statistics}
		\bvolume{7}(\bissue{4}),
		\bfpage{434}
		(\byear{1998})
	\end{barticle}
	\endbibitem
	
	\bibitem[\protect\citeauthoryear{Chatfield}{2003}]{chatfield2003analysis}
	\begin{bbook}
		\bauthor{\bsnm{Chatfield}, \binits{C.}}:
		\bbtitle{The Analysis of Time Series: an Introduction}.
		\bpublisher{Chapman and Hall/CRC}, \blocation{New York}
		(\byear{2003})
	\end{bbook}
	\endbibitem
	
	\bibitem[\protect\citeauthoryear{Gruber et~al.}{2011}]{2011A&A...533A..61G}
	\begin{barticle}
		\bauthor{\bsnm{Gruber}, \binits{D.}},
		\bauthor{\bsnm{Lachowicz}, \binits{P.}},
		\bauthor{\bsnm{Bissaldi}, \binits{E.}},
		\bauthor{\bsnm{Briggs}, \binits{M.S.}},
		\bauthor{\bsnm{Connaughton}, \binits{V.}},
		\bauthor{\bsnm{Greiner}, \binits{J.}},
		\bauthor{\bparticle{van~der} \bsnm{Horst}, \binits{A.J.}},
		\bauthor{\bsnm{Kanbach}, \binits{G.}},
		\bauthor{\bsnm{Rau}, \binits{A.}},
		\bauthor{\bsnm{Bhat}, \binits{P.N.}},
		\bauthor{\bsnm{Diehl}, \binits{R.}},
		\bauthor{\bparticle{von} \bsnm{Kienlin}, \binits{A.}},
		\bauthor{\bsnm{Kippen}, \binits{R.M.}},
		\bauthor{\bsnm{Meegan}, \binits{C.A.}},
		\bauthor{\bsnm{Paciesas}, \binits{W.S.}},
		\bauthor{\bsnm{Preece}, \binits{R.D.}},
		\bauthor{\bsnm{Wilson-Hodge}, \binits{C.}}:
		\bjtitle{Astronomy {\&} Astrophysics}
		\bvolume{533},
		\bfpage{61}
		(\byear{2011})
	\end{barticle}
	\endbibitem
	
	\bibitem[\protect\citeauthoryear{Hastings}{1970}]{1970Bimka..57...97H}
	\begin{barticle}
		\bauthor{\bsnm{Hastings}, \binits{W.K.}}:
		\bjtitle{Biometrika}
		\bvolume{57}(\bissue{1}),
		\bfpage{97}
		(\byear{1970})
	\end{barticle}
	\endbibitem
	
	\bibitem[\protect\citeauthoryear{Inglis et~al.}{2015}]{2015ApJ...798..108I}
	\begin{barticle}
		\bauthor{\bsnm{Inglis}, \binits{A.R.}},
		\bauthor{\bsnm{Ireland}, \binits{J.}},
		\bauthor{\bsnm{Dominique}, \binits{M.}}:
		\bjtitle{The Astrophysical Journal}
		\bvolume{798}(\bissue{2}),
		\bfpage{108}
		(\byear{2015})
	\end{barticle}
	\endbibitem
	
	\bibitem[\protect\citeauthoryear{Inglis et~al.}{2016}]{2016ApJ...833..284I}
	\begin{barticle}
		\bauthor{\bsnm{Inglis}, \binits{A.R.}},
		\bauthor{\bsnm{Ireland}, \binits{J.}},
		\bauthor{\bsnm{Dennis}, \binits{B.R.}},
		\bauthor{\bsnm{Hayes}, \binits{L.}},
		\bauthor{\bsnm{Gallagher}, \binits{P.}}:
		\bjtitle{The Astrophysical Journal}
		\bvolume{833}(\bissue{2}),
		\bfpage{284}
		(\byear{2016})
	\end{barticle}
	\endbibitem
	
	\bibitem[\protect\citeauthoryear{Li and Zhang}{2017}]{2017MNRAS.471L...6L}
	\begin{barticle}
		\bauthor{\bsnm{Li}, \binits{D.}},
		\bauthor{\bsnm{Zhang}, \binits{Q.M.}}:
		\bjtitle{Monthly Notices of the Royal Astronomical Society: Letters}
		\bvolume{471}(\bissue{1}),
		\bfpage{6}
		(\byear{2017})
	\end{barticle}
	\endbibitem
	
	\bibitem[\protect\citeauthoryear{Li et~al.}{2015}]{2015ApJ...813...59L}
	\begin{barticle}
		\bauthor{\bsnm{Li}, \binits{D.}},
		\bauthor{\bsnm{Ning}, \binits{Z.J.}},
		\bauthor{\bsnm{Zhang}, \binits{Q.M.}}:
		\bjtitle{The Astrophysical Journal}
		\bvolume{813}(\bissue{1}),
		\bfpage{59}
		(\byear{2015})
	\end{barticle}
	\endbibitem
	
	\bibitem[\protect\citeauthoryear{Li et~al.}{2017}]{2017A&A...597L...4L}
	\begin{barticle}
		\bauthor{\bsnm{Li}, \binits{D.}},
		\bauthor{\bsnm{Zhang}, \binits{Q.M.}},
		\bauthor{\bsnm{Huang}, \binits{Y.}},
		\bauthor{\bsnm{Ning}, \binits{Z.J.}},
		\bauthor{\bsnm{Su}, \binits{Y.N.}}:
		\bjtitle{Astronomy {\&} Astrophysics}
		\bvolume{597},
		\bfpage{4}
		(\byear{2017})
	\end{barticle}
	\endbibitem
	
	\bibitem[\protect\citeauthoryear{Li et~al.}{2018}]{2018ApJ...868L..33L}
	\begin{barticle}
		\bauthor{\bsnm{Li}, \binits{L.}},
		\bauthor{\bsnm{Zhang}, \binits{J.}},
		\bauthor{\bsnm{Peter}, \binits{H.}},
		\bauthor{\bsnm{Chitta}, \binits{L.P.}},
		\bauthor{\bsnm{Su}, \binits{J.}},
		\bauthor{\bsnm{Song}, \binits{H.}},
		\bauthor{\bsnm{Xia}, \binits{C.}},
		\bauthor{\bsnm{Hou}, \binits{Y.}}:
		\bjtitle{The Astrophysical Journal Letters}
		\bvolume{868}(\bissue{2}),
		\bfpage{33}
		(\byear{2018})
	\end{barticle}
	\endbibitem
	
	\bibitem[\protect\citeauthoryear{Metropolis et~al.}{1953}]{1953JChPh..21.1087M}
	\begin{barticle}
		\bauthor{\bsnm{Metropolis}, \binits{N.}},
		\bauthor{\bsnm{Rosenbluth}, \binits{A.W.}},
		\bauthor{\bsnm{Rosenbluth}, \binits{M.N.}},
		\bauthor{\bsnm{Teller}, \binits{A.H.}},
		\bauthor{\bsnm{Teller}, \binits{E.}}:
		\bjtitle{The Journal of Chemical Physics}
		\bvolume{21}(\bissue{6}),
		\bfpage{1087}
		(\byear{1953})
	\end{barticle}
	\endbibitem
	
	\bibitem[\protect\citeauthoryear{Ning}{2017}]{2017SoPh..292...11N}
	\begin{barticle}
		\bauthor{\bsnm{Ning}, \binits{Z.}}:
		\bjtitle{Solar Physics}
		\bvolume{292}(\bissue{1}),
		\bfpage{11}
		(\byear{2017})
	\end{barticle}
	\endbibitem
	
	\bibitem[\protect\citeauthoryear{Pesnell et~al.}{2012}]{2012SoPh..275....3P}
	\begin{barticle}
		\bauthor{\bsnm{Pesnell}, \binits{W.D.}},
		\bauthor{\bsnm{Thompson}, \binits{B.J.}},
		\bauthor{\bsnm{{Chamberlin, P. C.}}}:
		\bjtitle{Solar Physics}
		\bvolume{275}(\bissue{1}),
		\bfpage{3}
		(\byear{2012})
	\end{barticle}
	\endbibitem
	
	\bibitem[\protect\citeauthoryear{{Pugh} et~al.}{2017}]{2017A&A...602A..47P}
	\begin{barticle}
		\bauthor{\bsnm{{Pugh}}, \binits{C.E.}},
		\bauthor{\bsnm{{Broomhall}}, \binits{A.-M.}},
		\bauthor{\bsnm{{Nakariakov}}, \binits{V.M.}}:
		\bjtitle{\aap}
		\bvolume{602},
		\bfpage{47}
		(\byear{2017}).
		\arxivurl{1703.07294}.
		doi:\doiurl{10.1051/0004-6361/201730595}
	\end{barticle}
	\endbibitem
	
	\bibitem[\protect\citeauthoryear{Stangalini et~al.}{2012}]{2012A&A...539L...4S}
	\begin{barticle}
		\bauthor{\bsnm{Stangalini}, \binits{M.}},
		\bauthor{\bsnm{Giannattasio}, \binits{F.}},
		\bauthor{\bsnm{Del~Moro}, \binits{D.}},
		\bauthor{\bsnm{Berrilli}, \binits{F.}}:
		\bjtitle{Astronomy {\&} Astrophysics}
		\bvolume{539},
		\bfpage{4}
		(\byear{2012})
	\end{barticle}
	\endbibitem
	
	\bibitem[\protect\citeauthoryear{Su et~al.}{2016}]{2016ApJ...816...30S}
	\begin{barticle}
		\bauthor{\bsnm{Su}, \binits{J.T.}},
		\bauthor{\bsnm{Ji}, \binits{K.F.}},
		\bauthor{\bsnm{Banerjee}, \binits{D.}},
		\bauthor{\bsnm{Cao}, \binits{W.D.}},
		\bauthor{\bsnm{Priya}, \binits{T.G.}},
		\bauthor{\bsnm{Zhao}, \binits{J.S.}},
		\bauthor{\bsnm{Yu}, \binits{S.J.}},
		\bauthor{\bsnm{Ji}, \binits{H.S.}},
		\bauthor{\bsnm{Zhang}, \binits{M.}}:
		\bjtitle{The Astrophysical Journal}
		\bvolume{816}(\bissue{1}),
		\bfpage{30}
		(\byear{2016})
	\end{barticle}
	\endbibitem
	
	\bibitem[\protect\citeauthoryear{Tian et~al.}{2014}]{2014ApJ...786..137T}
	\begin{barticle}
		\bauthor{\bsnm{Tian}, \binits{H.}},
		\bauthor{\bsnm{DeLuca}, \binits{E.}},
		\bauthor{\bsnm{Reeves}, \binits{K.K.}},
		\bauthor{\bsnm{McKillop}, \binits{S.}},
		\bauthor{\bsnm{De~Pontieu}, \binits{B.}},
		\bauthor{\bsnm{Martinez-Sykora}, \binits{J.}},
		\bauthor{\bsnm{Carlsson}, \binits{M.}},
		\bauthor{\bsnm{Hansteen}, \binits{V.}},
		\bauthor{\bsnm{Kleint}, \binits{L.}},
		\bauthor{\bsnm{Cheung}, \binits{M.}},
		\bauthor{\bsnm{Golub}, \binits{L.}},
		\bauthor{\bsnm{Saar}, \binits{S.}},
		\bauthor{\bsnm{Testa}, \binits{P.}},
		\bauthor{\bsnm{Weber}, \binits{M.}},
		\bauthor{\bsnm{Lemen}, \binits{J.}},
		\bauthor{\bsnm{Title}, \binits{A.}},
		\bauthor{\bsnm{Boerner}, \binits{P.}},
		\bauthor{\bsnm{Hurlburt}, \binits{N.}},
		\bauthor{\bsnm{Tarbell}, \binits{T.D.}},
		\bauthor{\bsnm{Wuelser}, \binits{J.P.}},
		\bauthor{\bsnm{Kankelborg}, \binits{C.}},
		\bauthor{\bsnm{Jaeggli}, \binits{S.}},
		\bauthor{\bsnm{McIntosh}, \binits{S.W.}}:
		\bjtitle{The Astrophysical Journal}
		\bvolume{786}(\bissue{2}),
		\bfpage{137}
		(\byear{2014})
	\end{barticle}
	\endbibitem
	
	\bibitem[\protect\citeauthoryear{Vaughan}{2005}]{2005A&A...431..391V}
	\begin{barticle}
		\bauthor{\bsnm{Vaughan}, \binits{S.}}:
		\bjtitle{Astronomy {\&} Astrophysics}
		\bvolume{431},
		\bfpage{391}
		(\byear{2005})
	\end{barticle}
	\endbibitem
	
	\bibitem[\protect\citeauthoryear{Vaughan}{2010}]{2010MNRAS.402..307V}
	\begin{barticle}
		\bauthor{\bsnm{Vaughan}, \binits{S.}}:
		\bjtitle{Monthly Notices of the Royal Astronomical Society}
		\bvolume{402}(\bissue{1}),
		\bfpage{307}
		(\byear{2010})
	\end{barticle}
	\endbibitem
	
	\bibitem[\protect\citeauthoryear{Wang et~al.}{2018}]{2018ApJ...856L..16W}
	\begin{barticle}
		\bauthor{\bsnm{Wang}, \binits{F.}},
		\bauthor{\bsnm{Deng}, \binits{H.}},
		\bauthor{\bsnm{Li}, \binits{B.}},
		\bauthor{\bsnm{Feng}, \binits{S.}},
		\bauthor{\bsnm{Bai}, \binits{X.}},
		\bauthor{\bsnm{Deng}, \binits{L.}},
		\bauthor{\bsnm{Yang}, \binits{Y.}},
		\bauthor{\bsnm{Xue}, \binits{Z.}},
		\bauthor{\bsnm{Wang}, \binits{R.}}:
		\bjtitle{The Astrophysical Journal Letters}
		\bvolume{856}(\bissue{1}),
		\bfpage{16}
		(\byear{2018})
	\end{barticle}
	\endbibitem
	
	\bibitem[\protect\citeauthoryear{Wang et~al.}{2020}]{2020RAA....20....6W}
	\begin{barticle}
		\bauthor{\bsnm{Wang}, \binits{Z.-K.}},
		\bauthor{\bsnm{Feng}, \binits{S.}},
		\bauthor{\bsnm{Deng}, \binits{L.-H.}},
		\bauthor{\bsnm{Meng}, \binits{Y.}}:
		\bjtitle{Research in Astronomy and Astrophysics}
		\bvolume{20}(\bissue{1}),
		\bfpage{006}
		(\byear{2020})
	\end{barticle}
	\endbibitem
	
	\bibitem[\protect\citeauthoryear{Yuan et~al.}{2014}]{2014ApJ...792...41Y}
	\begin{barticle}
		\bauthor{\bsnm{Yuan}, \binits{D.}},
		\bauthor{\bsnm{Nakariakov}, \binits{V.M.}},
		\bauthor{\bsnm{Huang}, \binits{Z.}},
		\bauthor{\bsnm{Li}, \binits{B.}},
		\bauthor{\bsnm{Su}, \binits{J.}},
		\bauthor{\bsnm{Yan}, \binits{Y.}},
		\bauthor{\bsnm{Tan}, \binits{B.}}:
		\bjtitle{The Astrophysical Journal}
		\bvolume{792}(\bissue{1}),
		\bfpage{41}
		(\byear{2014})
	\end{barticle}
	\endbibitem
	
	\bibitem[\protect\citeauthoryear{Yuan et~al.}{2016}]{2016ApJS..224...30Y}
	\begin{barticle}
		\bauthor{\bsnm{Yuan}, \binits{D.}},
		\bauthor{\bsnm{Su}, \binits{J.}},
		\bauthor{\bsnm{Jiao}, \binits{F.}},
		\bauthor{\bsnm{Walsh}, \binits{R.W.}}:
		\bjtitle{The Astrophysical Journal Supplement Series}
		\bvolume{224}(\bissue{2}),
		\bfpage{30}
		(\byear{2016})
	\end{barticle}
	\endbibitem
	
	\bibitem[\protect\citeauthoryear{{Yuan} et~al.}{2019}]{2019ApJ...886L..25Y}
	\begin{barticle}
		\bauthor{\bsnm{{Yuan}}, \binits{D.}},
		\bauthor{\bsnm{{Feng}}, \binits{S.}},
		\bauthor{\bsnm{{Li}}, \binits{D.}},
		\bauthor{\bsnm{{Ning}}, \binits{Z.}},
		\bauthor{\bsnm{{Tan}}, \binits{B.}}:
		\bjtitle{\apjl}
		\bvolume{886}(\bissue{2}),
		\bfpage{25}
		(\byear{2019}).
		\arxivurl{1911.05217}.
		doi:\doiurl{10.3847/2041-8213/ab5648}
	\end{barticle}
	\endbibitem
	
\end{thebibliography}
\end{document}